\documentclass[showpacs,aps,prl,twocolumn]{revtex4-1}
\pdfoutput=1

\usepackage{cmap}
\usepackage{graphicx}
\usepackage{fncylab,environ}
\usepackage{url}

\usepackage{amssymb}
\usepackage{amsmath}

\usepackage{tikz}
\usepackage{pgfplots}

\usepackage{hyperref}

\usetikzlibrary{external}
\labelformat{figure}{Figure #1}
\labelformat{equation}{(#1)}
\labelformat{section}{Section #1}
\labelformat{table}{Table #1}

\makeatletter

\newcommand\figshapes{
\begin{figure}
    \centering
    \includegraphics[width=0.24\linewidth]{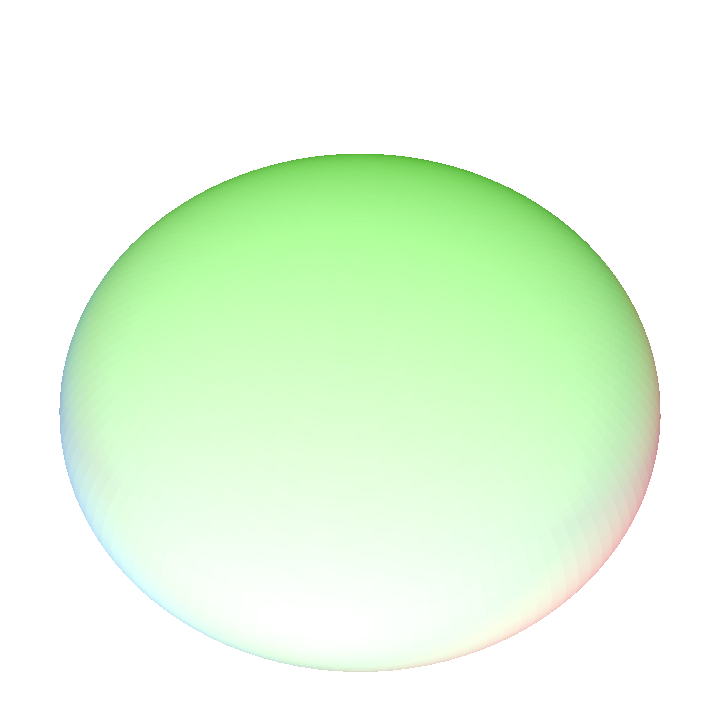}
    \includegraphics[width=0.24\linewidth]{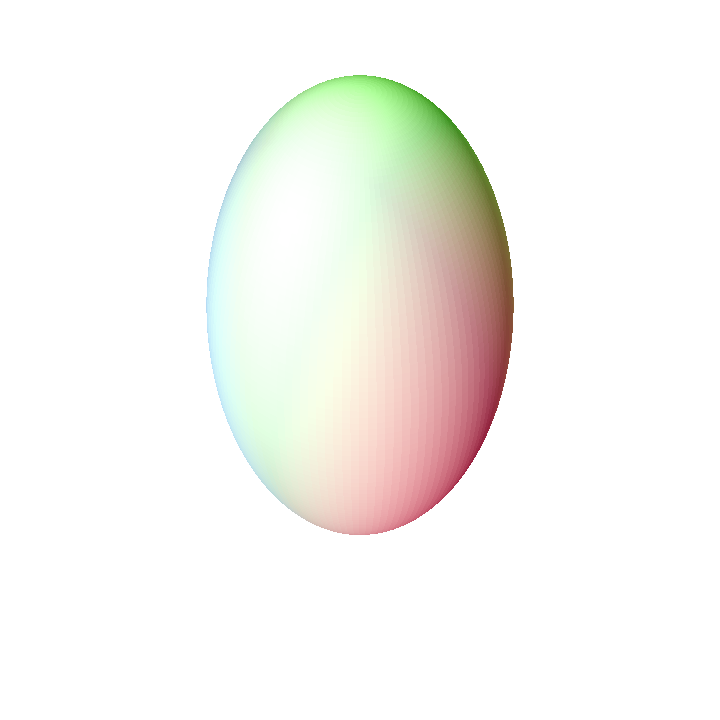}
    \includegraphics[width=0.24\linewidth]{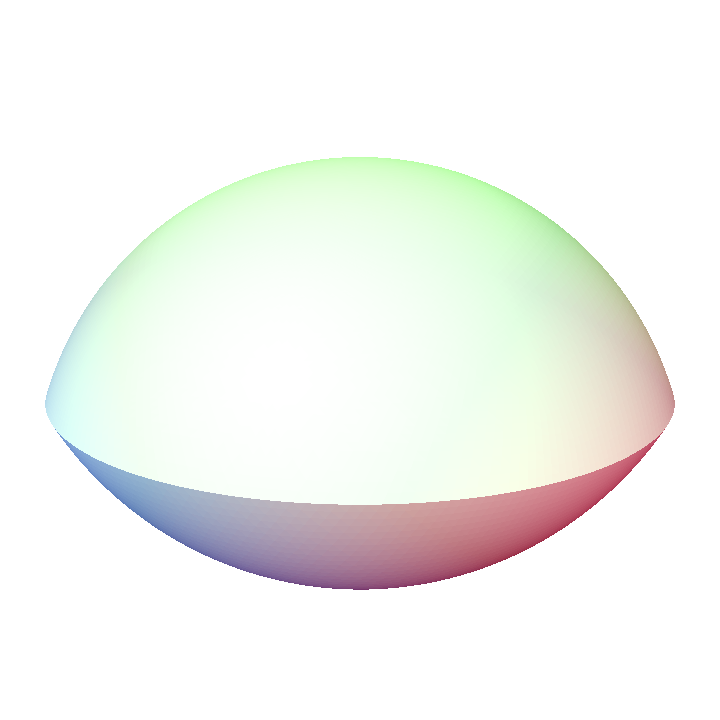}
    \includegraphics[width=0.24\linewidth]{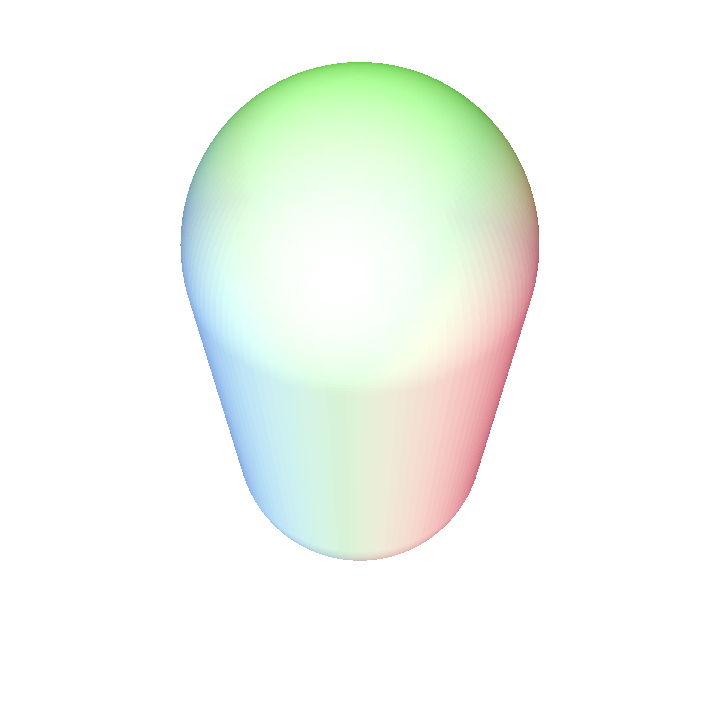}\\
    \includegraphics[width=0.24\linewidth]{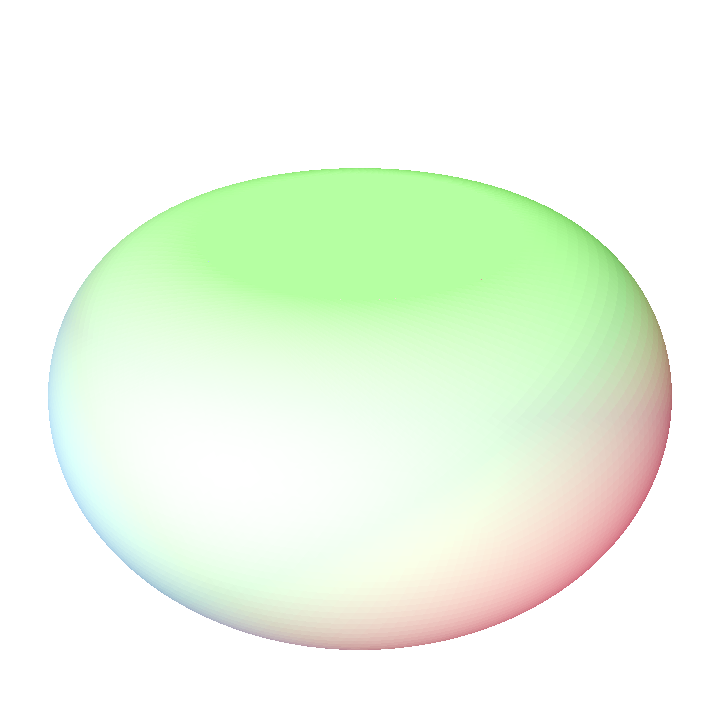}
    \includegraphics[width=0.24\linewidth]{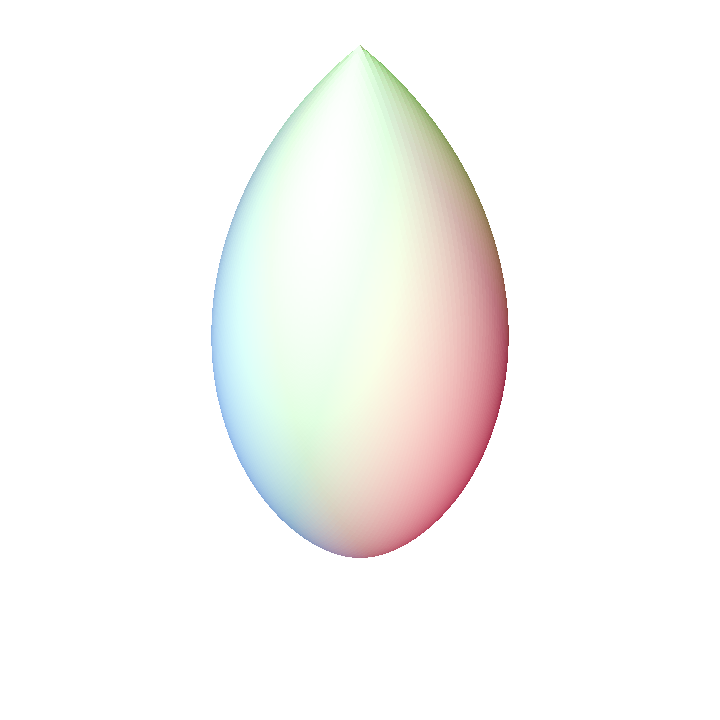}
    \includegraphics[width=0.24\linewidth]{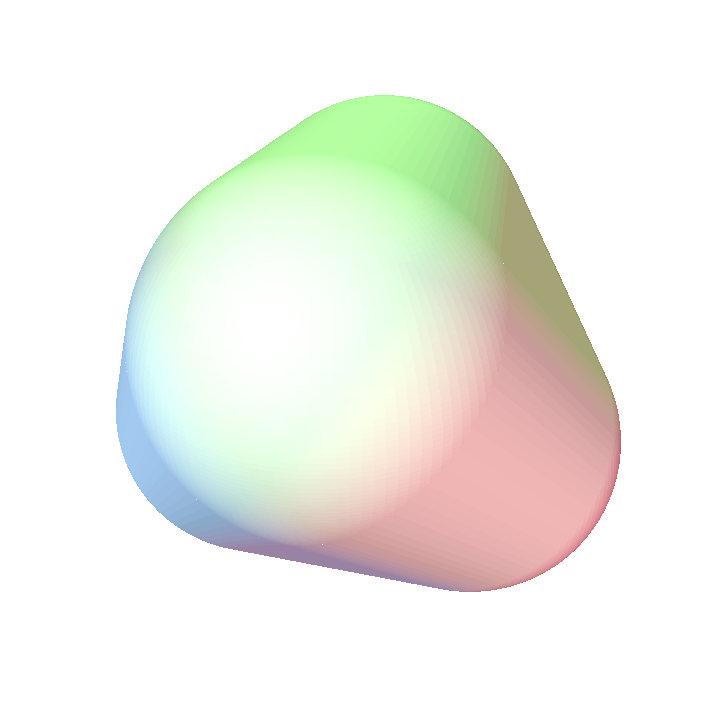}
    \includegraphics[width=0.24\linewidth]{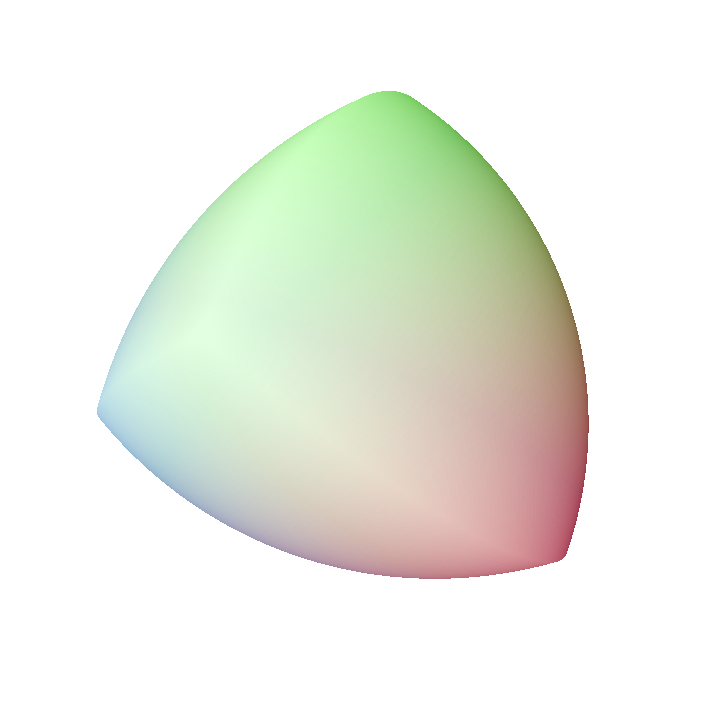}\\
    \includegraphics[width=0.24\linewidth]{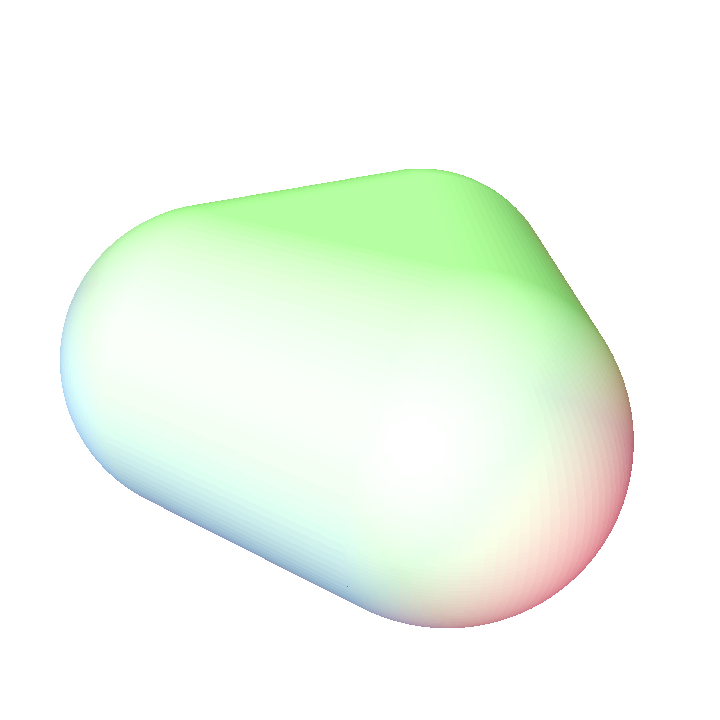}
    \includegraphics[width=0.24\linewidth]{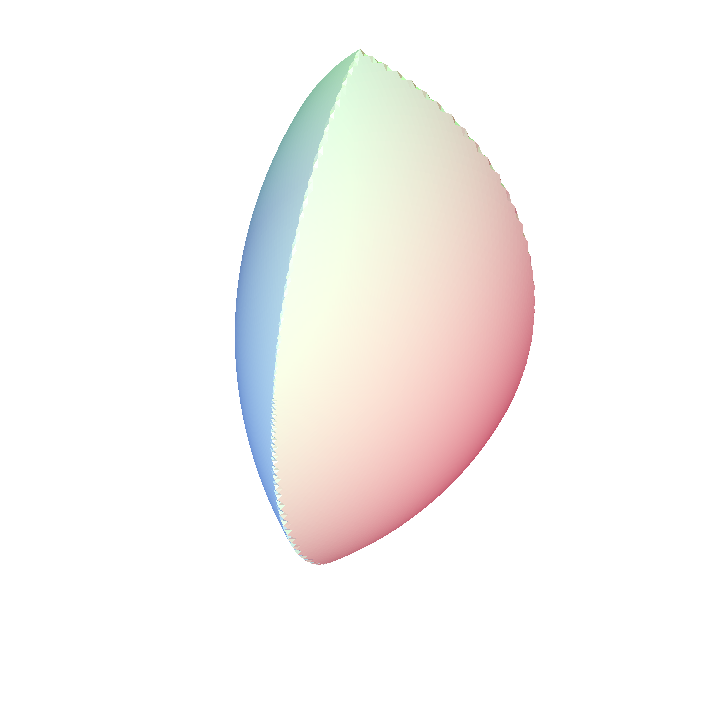}
    \includegraphics[width=0.24\linewidth]{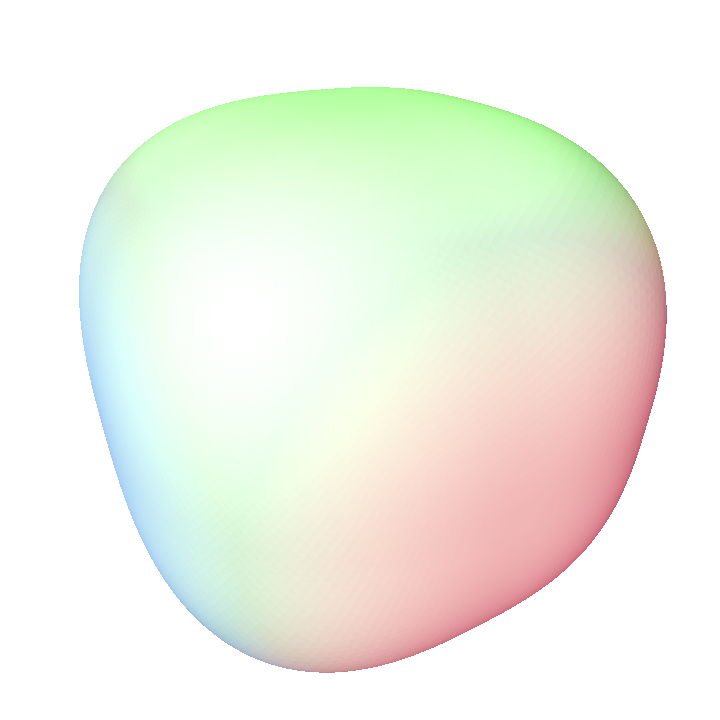}
    \includegraphics[width=0.24\linewidth]{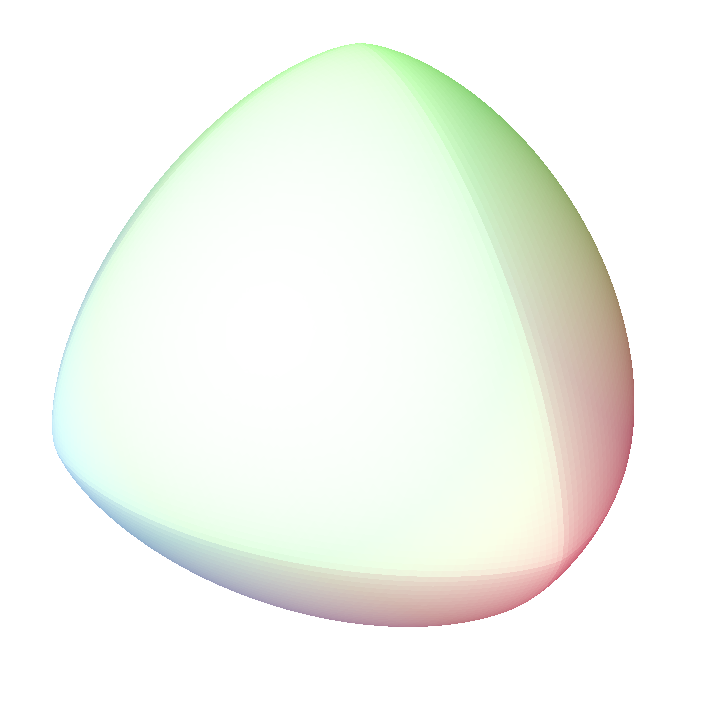}\\
    \caption{\label{fig:shapes}
    We calculate numerical slope estimates for 12 shape families, representatives of which are
    illustrated here in the order listed in \ref{tab:etas} (from top, left to right).
    }
\end{figure}
}
\newcommand\figell{
\begin{figure}
    \centering
    \tikzsetnextfilename{ell}
    \begin{tikzpicture}
	\begin{axis}[
		xlabel={aspect ratio $\alpha$}, ylabel={density $\phi$},
		xmin=1., xmax=1.3,
		ymin=0.64, ymax=0.72,
		domain=1:1.2,
		legend style={at={(0.8,0.97)},anchor=north east},
		legend entries={prolate,oblate},
	    ]
	    \addlegendimage{color=black, mark=diamond*, mark size=1.8pt,dashed};
	    \addlegendimage{color=black,mark=square*, mark size=1.2pt};
	    \addplot[color=black, only marks, mark=square*, mark size=1.2pt] table[x index=0, y index=1] {ell-pro.dat};
	    \addplot[color=black, only marks, mark=diamond*, mark size=1.4pt] table[x index=0,y index=1] {ell-obl.dat};
	    \addplot[color=black] {0.640 + 3.*0.202*(x-1.)};
	    \addplot[dashed,color=black] {0.640 + 3.*0.242*(x-1.)};
	\end{axis}
	\begin{axis}[
		width=0.5\linewidth,xshift=0.45\linewidth,yshift=0.03\linewidth,
		xlabel={\tiny shape parameter $p$}, ylabel={\tiny density $\phi$},
		xmin=0.84, xmax=1.4,
		ymin=0.63, ymax=0.72,
		domain=0.84:1.2,
		xticklabel style = { font=\tiny },
		xticklabel pos = right,
	        yticklabel style = { font=\tiny },
		every axis y label/.style={rotate=90,at={(-0.25,0.5)}},
		every axis x label/.style={at={(0.5,1.18)}},
	    ]
	    \addplot[color=black, only marks, mark=square*, mark size=1.2pt] table[x index=0, y index=1] {superballs-p.dat};
	    \addplot[color=black] {0.637 + 3.*0.142*(x-1.)};
	    \addplot[color=black] {0.637 + 3.*0.129*(1.-x)};
	\end{axis}
    \end{tikzpicture}
    \caption{\label{fig:ell}\label{fig:sup}
    Random packing density of prolate and oblate ellipsoids according to simulation results of Ref.\ \cite{Donev2007} together
    with lines illustrating predicted slopes at $\alpha=1$. Inset: Random packing density of octahedral and cubic superballs
    according to simulation results of Ref.\ \cite{Jiao2010} together
    with lines illustrating predicted slopes at $p=1$.
    }
\end{figure}
}
\newcommand\tabeta{
\begin{table}
    \begin{tabular}{c|c|c}
	shape family & $\eta$ & $\eta/(\overline{|\rho-\overline\rho|})$\\\hline
	oblate ellipsoid & $0.242\pm0.003$ & $0.94\pm0.01$\\
	prolate ellipsoid & $0.202\pm0.001$ & $0.79\pm0.01$\\
	lens & $0.216\pm0.004$ & $0.86\pm0.01$\\
	spherocylinder & $0.271\pm0.002$ & $1.08\pm0.01$\\
	spherodisk & $0.246\pm0.003$ & $1.36\pm0.02$\\
	spindle & $0.139\pm0.005$ & $0.77\pm0.03$\\
	rounded tetrahedron & $0.189\pm0.003$ & $1.45\pm0.02$\\
	tetrahedral puff & $0.138\pm0.005$ & $1.06\pm0.04$\\
	rounded triangle & $0.306\pm0.002$ & $1.31\pm0.01$\\
	triangular spindle & $0.235\pm0.004$ & $1.01\pm0.02$\\
	cubic superball & $0.142\pm0.001$ & $1.32\pm0.01$\\
	octahedral superball & $0.130\pm0.002$ & $1.20\pm0.02$\\
    \end{tabular}
    \caption{\label{tab:etas}Slopes of the random close packing density of one-parameter
    families of shapes at the point corresponding to the sphere. The slope parameter is defined as
    $\eta=(1/3)d\phi/d\alpha$ and depends on the parameterization. The normalized version,
    $\eta/(\overline{|\rho-\overline\rho|})$, where $\rho(\mathbf{u}) = dr_\alpha(\mathbf{u})/d\alpha$,
    is intrinsic to the shape family}
\end{table}
}

\begin{document}
\bibliographystyle{unsrt}

\title{The random packing density of nearly spherical particles}
\author{Yoav Kallus}
\affiliation{Santa Fe Institute, 1399 Hyde Park Road, Santa Fe, New Mexico 87501}

\date{\today}

\begin{abstract}
    Obtaining general relations between macroscopic properties of random assemblies, such as density, and the microscopic
    properties of their constituent particles, such as shape, is a foundational challenge in the study of amorphous materials.
    By leveraging existing understanding of the random packing of spherical particles, we estimate the
    random packing density for all sufficiently spherical shapes. Our method uses the ensemble of random packing configurations
    of spheres as a reference point for a perturbative calculation, which we carry to linear order in the deformation.
    A fully analytic calculation shows that all sufficiently spherical shapes
    pack more densely than spheres. Additionally, we use simulation data for spheres to calculate numerical estimates for
    nonspherical particles and compare these estimates to simulations.
\end{abstract}

\maketitle

Understanding the relationship between macroscopic properties of a random assembly and the microscopic properties, such
as shape and composition, of the particles that make it up is a fundamental problem in material science.
Colloidal suspensions \cite{Russel1992}, granular materials \cite{Mehta1994}, nanoparticle assemblies \cite{Valverde2006},
and biomaterials \cite{Ellis2001,Jiao2014} all feature disordered agglomerations of nonoverlapping particles.
When the particles are all undeformed spheres of the same size, the typical maximal density, measured by volume fraction, obtained by random
packing is $\phi\approx0.64$. Though the precise value of the density depends on preparation protocol,
the conventional value, to within a percent, has been reproduced in many experimental and
computational settings \cite{Ohern2002,Torquato2010,Berthier2011,Ozawa2012}.
This robust value suggests a purely geometric phenomenon, irrespective of particular
system-specific details such as residual interaction, hydrodynamics, or gravity, and this density is known as
the random-close-packing density.
The term random packing implies the lack of crystalline order in the packing, even at short ranges. Without this restriction, the largest volume fraction obtained
by packing of nonoverlapping spheres is $\pi/3\sqrt{2}\approx0.74$ \cite{Hales2015}.

While the case of equal-sized spheres is the canonical one,
applications in granular materials \cite{Abreu2003, Cho2006}, biophysics \cite{Katzav2006,Wong2003},
liquid crystals \cite{Shah2012}, and self-assembly \cite{Glotzer2007, Damasceno2012}
have driven interest in the random packing of nonspherical particles, as well as flexible particles \cite{Laso2009}
and polydisperse samples \cite{Jain2013}.
Simulations and experimental work have focused on particular shapes such as ellipsoids \cite{Donev2004,Donev2007}, spherocylinders \cite{Williams2003,Zhao2012},
and Platonic polyhedra \cite{Jiao2011,HajiAkbari2009}.
Convex shapes appear to achieve random packing densities well above that of spheres. This situation is similar to the situation
for the optimal (nonrandom) packing of nonspherical particles: a conjecture, attributed to Ulam, posits that
spheres have the lowest optimal packing density among all convex shapes \cite{Gardner1995}. Simulations with a large collection
of shapes have verified in each case that the shape can indeed be packed more densely than spheres \cite{Graaf2011}, and
it is proven that the same is true for any sufficiently spherical shape, where asphericity is measured by $\gamma$, the radius ratio
of smallest circumscribed sphere to largest inscribed sphere \cite{Kallus2014,Kallus2015}. It has also been conjectured that
all convex shapes below a certain asphericity (Jiao and Torquato suggest $\gamma<1.2$) have random packing densities above
that of spheres \cite{Jiao2011,Baule2013}. A mean-field calculation, focusing on axis-symmetric shapes, has
provided some theoretical backing to the conjecture \cite{Baule2013,Baule2014}.
In contrast to Ulam's conjecture, the sphere cannot be a global minimum,
since very elongated particles have random packing densities in inverse proportion to their aspect ratio,
so they can be arbitrarily small \cite{Philipse1996}.

Moreover, unlike the optimal density, which is rigorously defined, the random-close-packing density does not have an accepted mathematical definition.
Operationally, random close packing is often defined as the packing obtained through some fixed numerical protocol \cite{Zong2014}, for example,
a molecular dynamics simulation of slowly expanding hard particles \cite{Torquato2000} or a sequence of energy minimization of a system of
soft particles to converge to the density where the repulsion
energy is on the verge of becoming nonzero \cite{Ohern2005}. We present a heuristic calculation that predicts,
irrespective of the preparation protocol, that any sufficiently spherical convex shape will yield a higher density than spheres would
given the same protocol. An important assumption we make about the protocol is that it produces isostatic packings when
applied to spherical particles; that is, the number of interparticle contacts is equal to the number of degrees of freedom,
as predicted by Maxwellian constraint counting (when unconstrained ``rattlers'' are removed). Isostaticity is observed in nearly
all random packing protocols for spheres, but random packings of nonspherical particles are usually hypostatic, with far fewer
contacts than degrees of freedom \cite{Donev2007}. 

The calculation we present is an extension to general convex shapes in three dimensions 
of the calculation performed by Donev et al.\ in two dimensions for ellipses \cite{Donev2007}.
The calculation relies on the assumption that for nearly spherical particles, the following two
procedures would produce similar results: (1) applying the random packing protocol to spheres
and then deforming the spheres to the shape of interest while maintaining a compressive pressure; 
and (2) applying the random packing protocol directly to the particles of interest.
The rationale is, that nearly spherical particles will behave like sphere for the majority of the
compression protocol, and their asphericity will only come into effect near the end, when the configuration
is similar to a random packed configuration of spheres. This argument has been shown to hold true
in the case of ellipses \cite{Donev2007}.

Therefore, we begin with a random packed configuration of $N$ spheres of equal diameter $\sigma$, and we continuously deform it into a packing of
particles all congruent to a given body. Geometrically, the body is described as a compact subset $K\subseteq \mathbb{R}^3$.
We can also define its radial function $r_K(\mathbf{u}) = \max_{\lambda\mathbf{u}\in K}\lambda$ as the distance
from the center of $K$ to its boundary along a certain direction $\mathbf{u}$.
Without loss of generality, we will assume that the largest ball contained in $K$ is of diameter $\sigma$, and let $\sigma+\Delta\sigma$
be the diameter of the smallest concentric ball that contains $K$. Thus, $\epsilon=\Delta\sigma/\sigma=\gamma-1$ measures the eccentricity
of $K$. We consider a one-parameter family of nested convex shapes $K_t$, $0\le t\le1$ such that $K_0$ is the ball of diameter $\sigma$
and $K_1=K$. As a first step, we calculate the change in volume if
the orientation of each particle is fixed: we assign to each sphere a rotation matrix $R_i$, $i=1,\ldots,N$, so that the packing at time $t$
is given by the collection $\{ R_i K_t + \mathbf{x}_i(t) : i=1,\ldots,N\}$. Since the sphere is invariant
under rotations, any choice of the matrices $R_i$ gives the same configuration at $t=0$.

Throughout the deformation, we maintain a compressive pressure on the sample by applying external forces
to the particles along the boundary. We label by $\mathbf{F}_i$ the external force applied to particle $i$
($\mathbf{F}_i=0$ for particles away from the boundary).
By definition, we measure the change in volume by the work performed by
the particles against these forces: $p\Delta V = \sum_{i=0}^N \langle \mathbf{F}_i, \mathbf{x}_i(1)-\mathbf{x}_i(0)\rangle$.
From the balance of forces on each particle at $t=0$ we have that
\begin{equation}\label{eq:force-balance}
    \mathbf{F}_i = \sum_{j\in\partial i} f_{ij} \mathbf{n}_{ij}\text,
\end{equation}
where $\partial i$ is the set of indices of particles in contact with particle $i$ at time $t=0$,
$\mathbf{n}_{ij}$ is a unit normal vector at the contact point between particles
$i$ and $j$, and $f_{ij}=f_{ji}$ is the contact force magnitude.
The pairs of particles that are in contact, their contact normal, and contact force
will evolve throughout the deformation, but we use the symbols $\partial$, $\mathbf{n}_{ij}$,
and $f_{ij}$ without a time argument to denote their values at $t=0$.
We also use $\sum_{i\sim j}$ to denote a sum over unordered pairs $\{i,j\}$ such that $j\in\partial i$.
Summing over all particles, we obtain
\begin{equation}\label{eq:force-balance2}
    p\Delta V = \sum_{i\sim j} f_{ij} \langle\mathbf{n}_{ij},\Delta\mathbf{x}_i-\Delta\mathbf{x}_j\rangle\text.
\end{equation}
The pressure $p$ is an arbitrary conversion factor between energy and volume
dimensions, which we fix by applying \ref{eq:force-balance2} to an infinitesimal uniform expansion,
$\Delta\mathbf{x}_i = \alpha \mathbf{x}_i$ and $\Delta V = 3\alpha V$, to
obtain $3pV = \sigma \sum_{i\sim j} f_{ij}$, or equivalently, 
\begin{equation}\label{eq:meanforce}
    \langle f \rangle = \frac{6pV}{N\sigma \langle|\partial i|\rangle_i}\text,
\end{equation}
where $\langle|\partial i|\rangle_i$ is the mean coordination number, which is $6$ for isostatic
packings.

We first consider pairs of particles that are at contact at time $t=0$ and remain so
throughout the deformation. Their relative displacement satisfies
\begin{equation}\label{eq:contact}
    \Delta\mathbf{x}_i-\Delta\mathbf{x}_j = r_K(R_i^{-1} \mathbf{u}_{ij}) \mathbf{u}_{ij} - r_K(R_j^{-1} \mathbf{u}_{ji}) \mathbf{u}_{ji} - \sigma \mathbf{n}_{ij}\text,
\end{equation}
where $\mathbf{u}_{ij}$ ($\mathbf{u}_{ji}$) is a unit vector in the direction from the center of the particle $i$ ($j$)
to the point of contact at time $t=1$.
We assume the change in the packing geometry is continuous, so that $\mathbf{u}_{ij}$ and $\mathbf{u}_{ji}$ approach $\pm\mathbf{n}_{ij}$ in the
limit that $\epsilon\to 0$. Therefore, as a zeroth-order approximation, we plug in $\mathbf{u}_{ij}=-\mathbf{u}_{ji}=\mathbf{n}_{ij}$ to
get $\Delta\mathbf{x}_i-\Delta\mathbf{x}_j = \left[r_K(R_i^{-1} \mathbf{n}_{ij}) + r_K(-R_j^{-1} \mathbf{n}_{ij}) - \sigma\right] \mathbf{n}_{ij}$.
We estimate the error between the zeroth order approximation and the actual displacement by assuming self-consistently that
the size of the displacement does not exceed the order $O(\sigma\epsilon)$ predicted by the approximation, and therefore,
that $\mathbf{u}_{ij}-\mathbf{n}_{ij} = O(\epsilon)$ as well. 
Together with the fact that the function $r_K(\mathbf{u})$ is $O(\sigma\epsilon^{1/2})$-Lipschitz continuous \cite{convex-lipschitz},
we obtain
\begin{equation}\label{eq:dr}
    \begin{aligned}
    \langle\mathbf{n}_{ij},&\mathbf{\Delta x}_i-\mathbf{\Delta x}_j\rangle = \\&\Delta r(R_i^{-1} \mathbf{n}_{ij}) + \Delta r(-R_j^{-1} \mathbf{n}_{ij})+
    O(\sigma\epsilon^{3/2})\text,
    \end{aligned}
\end{equation}
where $\Delta r (\mathbf{u}) = r_K(\mathbf{u}) - (\sigma/2)$.

For contacts that are broken during the deformation, the right hand side of \ref{eq:dr} serves
only as a lower bound, but we assume that the displacement is still of order $O(\sigma \epsilon)$.
Since the initial number of contacts is the minimal number required to maintain stability,
the number of contacts broken is at most equal to the number of new contacts made \cite{Donev2007}. A rough upper bound for the 
number of new contacts made is the number of pairs of spheres, not initially in contact, which would
come to overlap if we dilated each sphere by a factor $1+\epsilon$. The number of such pairs
asymptotically for small $\epsilon$ is
known in random packing configurations of spheres to approach $A N\epsilon^{\sim 0.6}$ \cite{Charbonneau2012}.
Assuming that the forces associated with broken contacts
are at most of average magnitude (typically contacts with smaller forces will break first), we
see that the contribution to $\Delta V$ from broken contacts in \ref{eq:dr} is of order at most
$N\epsilon^{\sim 1.6}$. Thus, to leading order in $\epsilon$, 
\begin{equation}\label{eq:chvol2}
    p\Delta V = \sum_{i}\sum_{j\in\partial i} f_{ij} \Delta r(R_i^{-1} \mathbf{n}_{ij}) + O(V\epsilon^{3/2})\text.
\end{equation}

We now consider what the change in volume would be if instead of holding the particle
orientations fixed, we allowed them to rotate freely. As the system is under a compressive
pressure, it will tend to adopt the orientations that allow it to minimize the volume.
Therefore we assume that the change in volume will be the same as if we held the particle orientations fixed during
the deformation but rotated them in
advance (while they are spheres, and their orientations is physically irrelevant) to the orientations that
will yield the lowest volume at the end of the deformation. Namely, we assume that $p\Delta V$
is equal to the minimum of the right hand side of \ref{eq:chvol2} over all choices of the
rotation matrices $R_i$. The error term can be moved outside of the minimization, and we are left with
\begin{equation}\label{eq:chvol3}
    p\Delta V = \min_{\substack{R_i\in SO(3)\\\text{for all }i}}\left[\sum_{i}\sum_{j\in\partial i} f_{ij} \Delta r(R_i^{-1} \mathbf{n}_{ij})\right] + O(V\epsilon^{3/2})\text.
\end{equation}
Remarkably, each term in the outer sum being minimized depends only on the orientation of a single particle
and can be minimized independently of all the other terms. Also, as $R_i$ is a dummy variable,
we can replace it with its inverse, as we do for neatness henceforth.

The density of the final packing is given by $Nv_K/V(1)$, where $v_K$ is the volume of a single particle and is given by
\begin{equation}
    \begin{aligned}
	v_K &= \int_{\mathbf{u}\in S^2} r_K(\mathbf{u})^3 d^2\mathbf{u} \\
	&= \tfrac{4\pi}3 (\sigma/2)^3 + 4\pi(\sigma/2)^2\overline{\Delta r} + O(\sigma^3\epsilon^2)\text,
    \end{aligned}
\end{equation}
where $\overline{\cdot}$ represents the average over the unit sphere $S^2$. Rewriting \ref{eq:chvol3} in terms of the intensive
density instead of the extensive volume and eliminating the arbitrary pressure by using \ref{eq:meanforce}, we get
\begin{equation}\label{eq:density}
    \frac{\Delta\phi}{3\phi_0} =
    \frac{\overline{\Delta r(\mathbf{u})}}{\sigma/2} -
    \frac{\left\langle \displaystyle{\min_R \sum_{j\in \partial i}} f_{ij} \Delta r(R \mathbf{n}_{ij})\right\rangle_i}{(\sigma/2)\langle |\partial i|\rangle_i \langle f\rangle} +
    O(\epsilon^{3/2})\text,
\end{equation}
where the average $\langle\cdot\rangle_i$ represents an average over the particles in the initial packing.
The calculation is valid also in dimensions $d\neq3$, in which case $d$ should be substituted for $3$ in \ref{eq:density}.

If, instead of minimizing over $R$ in \ref{eq:density}, we averaged over $R$,
the first two terms on the right hand side would cancel. Since the mean
serves an upper bound for the minimum, this argument immediately gives
$\Delta\phi > - O(\epsilon^{3/2})$. This is almost the claim we wish to make,
namely, that the random packing density of any nearly spherical convex shape is
larger than that of spheres.
To strictly bound $\Delta\phi$ above 0, we have to find a better lower bound on the
gap between the minimum and the average.

Consider a particle $i$, then we are interested in the minimum of
$g_i(R) = \sum_{j\in \partial i} f_{ij} \Delta r(R \mathbf{n}_{ij})$. We can average
$g_i(R)$ over all rotations that map a fixed point, say the north pole $\mathbf{z}$, on $S^2$ to a given point $\mathbf{v}$
to obtain a function over $S^2$:
\begin{equation}
    h_i(\mathbf{v}) = \tfrac{1}{2\pi} \int_{\theta=0}^{2\pi} d\theta
    \sum_{j\in \partial i} f_{ij} \Delta r(R_\mathbf{v} R_z(\theta) \mathbf{n}_{ij})\text,
\end{equation}
where $R_\mathbf{v}$ is the rotation mapping $\mathbf{z}$ to $\mathbf{v}$ around
the axis perpendicular to both, and $R_z(\theta)$ is the rotation about the $z$-axis
by an angle of $\theta$. Clearly, the minimum of $h_i(\mathbf{v})$ is no smaller than
the minimum of $g_i(R)$. The structure of the linear operator
$\Phi_i\colon \Delta r(\mathbf{u})\mapsto h_i(\mathbf{v})$ is that of a convolution with
a zonal (i.e.\ invariant over rotations that fix the pole) measure over $S^2$ \cite{convolutions}. This structure
allows us to bound the $L_1$ norm of the deviation of $h_i(\mathbf{v})$ from its mean in terms
of that deviation in $\Delta r$. This procedure also gives the following bound on the minimum:
\begin{equation}\label{eq:minL1}
    \begin{aligned}
	\min_R g_i(R) - \overline{g_i(R)} &\le \min_\mathbf{v} h_i(\mathbf{v}) - \overline{h_i(\mathbf{v})}\\
	& \le c_i \overline{|\Delta r(\mathbf{u})-\overline{\Delta r(\mathbf{u})}|}\text.
    \end{aligned}
\end{equation}
We do not give a complete description of this procedure, as it is essentially the same as Section 4
of \cite{Kallus2014}. The constant $c_i$ is strictly positive whenever the spherical harmonic expansion of the zonal
measure used in the convolution has no terms that vanish. Therefore, as this will not happen generically
(not to mention to a fraction of particles approaching one), we can safely assume that the average value
$\langle c_i\rangle_i$ is strictly positive. Using the bound \ref{eq:minL1} in \ref{eq:density},
we immediately get that $\Delta\phi > c \overline{|\Delta r(\mathbf{u})-\overline{\Delta r(\mathbf{u})}|}$
for some constant $c$.
As the right hand side is zero only for spheres, we have obtained the result we were after.

Having obtained the theoretical result for general shapes and general
packing protocols, we now wish to calculate numerical estimates for specific shapes and protocols.
Consider a family of convex shapes, parameterized by some single variable, $\alpha$, that includes the sphere
of diameter $\sigma$ at $\alpha=\alpha_0$. Let $r_\alpha(\mathbf{u})$ be the radial function describing the
shapes, and let $\rho(\mathbf{u}) = (dr_\alpha(\mathbf{u})/d\alpha|_{\alpha_0^+})/(\sigma/2)$.
We define $\eta$ as a measure of the slope of the random packing density subject to a given protocol
as a function of the parameter $\alpha$:
\begin{equation}\label{eq:eta-def}
    \eta = \left.\frac1{3\phi}\frac{d\phi(\alpha)}{d\alpha}\right|_{\alpha_0^+}\text.
\end{equation}
We take the derivative in the direction of positive $\alpha$, as $\phi(\alpha)$ is
usually not smooth at $\alpha_0$. From \ref{eq:density} we have
\begin{equation}\label{eq:slope}
    \eta =
    \overline{\rho(\mathbf{u})} -
    \frac1{\langle |\partial i|\rangle_i \langle f\rangle}\left\langle \min_R \sum_{j=\partial i} f_{ij} \rho(R \mathbf{u}_{ij})\right\rangle_i\text,
\end{equation}
where $\langle \cdot\rangle_i$ denotes the average over particles in random packing configuration of spheres
produced by the same protocol.

We estimate $\eta$ for a few examples of shape families in the case
of the random packing protocol of Jin and Makse \cite{Jin2010}.
To numerically calculate \ref{eq:slope}, we use
a dataset that includes the coordinates and forces of a configuration of $10^4$
spheres in a periodic cubic box \cite{Jin2010}.


\figshapes

Consider first the family of prolate ellipsoid, parameterized by aspect ratio.
The infinitesimal deformation is $\rho(\theta,\varphi) = \cos^2\theta$ and
we can solve the minimization analytically:
\begin{equation}\label{eq:slope-prol}
    \eta_\text{prol}=
    \frac13 -
    \frac1{\langle |\partial i|\rangle_i\langle f\rangle}\left\langle \lambda_{\min}(F_i)\right\rangle_i\text,
\end{equation}
where $F_i=\sum_{j\in\partial i} f_{ij} \mathbf{n}_{ij}\otimes\mathbf{n}_{ij}$ is a symmetric
tensor describing the stress on particle $i$, and $\lambda_{\min}$ is
its smallest eigenvalue. We calculate a numerical value of $\eta_\text{prol} = 0.202\pm0.001$.
The error estimate quoted includes only statistical error. There could be systematic error from
the fact that the dataset we use is not precisely at the isostatic point and from finite-size effects.

\tabeta

For an oblate ellipsoid, $\rho=1-\cos^2\theta$, and so
\begin{equation}\label{eq:slope-obl}
	\eta_\text{obl}
        =\frac1{\langle |\partial i|\rangle_i\langle f\rangle}\left\langle \lambda_{\max}(F)\right\rangle_i - \frac13\text.
\end{equation}
Numerically, we get $\eta_\text{obl} = 0.242\pm0.003$.
In the case of triaxial ellipsoids, we consider for each $0<\mu<1$ the family where the
principal axes are given by $\sigma<\alpha^{\mu}\sigma<\alpha\sigma$.
Then $\rho(\theta,\phi) = \cos^2\theta+\mu\sin^2\theta\cos^2\phi$, and we get
$\eta_\text{triax}=(1-\mu)\eta_\text{prol}+\mu\,\eta_\text{obl}$.

Leaving aside ellipsoids, where minimization can
be done analytically, we consider other shape families of interest, illustrated in
\ref{fig:shapes} and defined in the appendix. 
We numerically find $R_i$ that minimizes $\sum_{j\in\partial i}f_{ij}\rho(R_i \mathbf{n}_{ij})$
for each of the $10^4$ spheres in the dataset and for each shape family.
The resulting numerical values for the slope $\eta=(1/3) d\phi/d\alpha$ are given
in \ref{tab:etas}. Since the value of $\eta$ depends on the way
the family of shapes is parameterized, we also give the normalized value
$\eta/\overline{|\rho-\overline\rho|}$
\footnote{The preprint source files include replication code for this
calculation.}.

\figell

Rounded tetrahedra appear to give the largest normalized improvement
in packing density out of all the families considered. This
observation is in harmony with the fact that regular tetrahedra, out
of all convex shapes that have been studied, seem to have the largest
random close packing density \cite{Jaoshvili2010}.

We compare our predicted slopes to simulation data for prolate and oblate ellipsoids \cite{Donev2007}
and for superballs \cite{Jiao2010} in \ref{fig:sup}.
The protocols used in these simulations are differenent than the protocol used
to obtain the data on which we based our numerical calculation, so
this comparison should be interpreted with caution.
It is hard to tell from the limited data
if our calculation systematically overestimates the slope of the density curve,
or whether nonlinearities of the curve quickly cause the data to diverge from the
linear estimate. To resolve this uncertainty, data for shapes in these families
closer to the spheres will be needed.

The method we present allows us to extend the existing robust knowledge about the random packing
behavior of spheres to all sufficiently spherical shapes. We perform some numerical
calculations for specific one- and two-parameter families of three-dimensional shapes,
but our method is applicable to any nearly spherical shape in any number of dimensions.
As such, it provides a valuable tool to test general ideas about the random packing
behavior of nonspherical shapes, such as the conjecture that spheres minimize this
density among all sufficiently spherical convex shapes. Our calculation predicts that indeed spheres are a local
minimum, providing substantial backing to the conjecture. It is worth noting that this
claim holds irrespective of the number of dimensions. In this aspect, the optimal (nonrandom)
packing problem is completely different: the $d$-dimensional sphere in any dimension
other than 3 does not appear to be even a local minimum of the optimal packing density \cite{Kallus2014}.

\textbf{Acknowledgments}. This work was supported by the Santa Fe Institute Omidyar Fellowship.
H.\ Makse and A.\ Baule were helpful in suggesting simulation data sources and discussing the analysis and
results.

\bibliography{rcpulam}

\begin{thebibliography}{10}

\bibitem{Russel1992}
W.~B. Russel, D.~A. Saville, and W.~R. Schowalter.
\newblock {\em Colloidal Dispersions}.
\newblock Cambridge Univ Press, 1992.

\bibitem{Mehta1994}
A.~Mehta, editor.
\newblock {\em Granular matter: an interdisciplinary approach}.
\newblock Springer-Verlag, New York, 1994.

\bibitem{Valverde2006}
J.~M. Valverde and A.~Castellanos.
\newblock {\em Europhys. Lett.}, 75(6):985, 2006.

\bibitem{Ellis2001}
R.~J. Ellis.
\newblock {\em Trends Biochem. Sci.}, 26:597--604, 2001.

\bibitem{Jiao2014}
Y.~Jiao, T.~Lau, H.~Hatzikirou, M.~Meyer-Hermann, J.~C. Corbo, and S.~Torquato.
\newblock {\em Phys. Rev. E}, 89:022721, 2014.

\bibitem{Ohern2002}
C.~S. O'Hern, S.~A. Langer, A.~J. Liu, and S.~R. Nagel.
\newblock Random packings of frictionless particles.
\newblock {\em Phys. Rev. Lett.}, 88(7), 2002.

\bibitem{Torquato2010}
S.~Torquato and Stillinger~F. H.
\newblock {\em Rev. Mod. Phys.}, 82:2633, 2010.

\bibitem{Berthier2011}
L.~Berthier, G.~Biroli, J.-P. Bouchaud, L.~Cipelletti, and W.~van Saarloos,
  editors.
\newblock {\em Dynamical Heterogeneities in Glasses, Colloids, and Granular
  Media}.
\newblock Oxford Univ. Press, Oxford, 2011.

\bibitem{Ozawa2012}
M.~Ozawa, T.~Kuroiwa, T.~A. Ikeda, and K.~Miyazaki.
\newblock {\em Phys. Rev. Lett.}, 109:205701, 2012.

\bibitem{Hales2015}
T.~Hales, M.~Adams, G.~Bauer, D.~T. Dang, J.~Harrison, T.~L. Hoang,
  C.~Kaliszyk, V.~Magron, S.~McLaughlin, T.~T. Nguyen, T.~Q. Nguyen, T.~Nipkow,
  S.~Obua, J.~Pleso, Jason R., Alexey S., A.~H.~T. Ta, T.~N. Tran, D.~T. Trieu,
  J.~Urban, K.~K. Vu, and R.~Zumkeller.
\newblock 2015.
\newblock arXiv:1501.02155.

\bibitem{Abreu2003}
C.~R.~A. Abreu, F.~W. Tavares, and M.~Castier.
\newblock {\em Powder Tech.}, 134:167, 2003.

\bibitem{Cho2006}
G.~Cho, J.~Dodds, and J.~Santamarina.
\newblock {\em J. Geotech. Geoenviron. Eng.}, 132(5):591--602, 2006.

\bibitem{Katzav2006}
E.~Katzav, M.~Adda-Bedia, and A.~Boudaoud.
\newblock {\em Proc. Nat. Acad. Sci. USA}, 103(50):18900--18904, 2006.

\bibitem{Wong2003}
G.~C.~L. Wong, A.~Lin, Tang~J. X., Y.~Li, P.~A. Janmey, and C.~R. Safinya.
\newblock {\em Phys. Rev. Lett.}, 91:018103, 2003.

\bibitem{Shah2012}
A.~A. Shah, H.~Kang, K.~L. Kohlstedt, K.~H. Ahn, S.~C. Glotzer, C.~W. Monroe,
  and M.~J. Solomon.
\newblock {\em Small}, 8(10):1551--1562, 2012.

\bibitem{Glotzer2007}
S.~C. Glotzer and M.~J. Solomon.
\newblock {\em Nature Materials}, 6:557, 2007.

\bibitem{Damasceno2012}
P.~Damasceno, M.~Engel, and S.~C. Glotzer.
\newblock {\em Science}, 337:453, 2012.

\bibitem{Laso2009}
M.~Laso, N.~Ch. Karayiannis, K.~Foteinopoulou, M.~L. Mansfield, and M.~Kroger.
\newblock {\em Soft Matter}, 5(9):1762--1770, 2009.

\bibitem{Jain2013}
A.~Jain, M.~J. Metzger, and B.~J. Glasser.
\newblock {\em Powder Tech.}, 237:543--553, 2013.

\bibitem{Donev2004}
A.~Donev, I.~Cisse, D.~Sachs, E.~A. Variano, F.~H. Stillinger, R.~Connelly,
  S.~Torquato, and P.~M. Chaikin.
\newblock {\em Science}, 303(5660):990–993, 2004.

\bibitem{Donev2007}
A.~Donev, R.~Connelly, F.~H. Stillinger, and S.~Torquato.
\newblock {\em Phys. Rev. E}, 75(5):051304, 2007.

\bibitem{Williams2003}
S.~R. Williams and A.~P. Philipse.
\newblock {\em Phys. Rev. E}, 67:051301, 2003.

\bibitem{Zhao2012}
J.~Zhao, S.~Li, E.~Zou, and A.~Yu.
\newblock {\em Soft Matter}, 8:1003--1009, 2012.

\bibitem{Jiao2011}
Y.~Jiao and S.~Torquato.
\newblock {\em Phys. Rev. E}, 84:041309, 2011.

\bibitem{HajiAkbari2009}
A.~Haji-Akbari, M.~Engel, A.~S. Keys, Z.~Xiaoyu, R.~G. Petschek,
  P.~Palffy-Muhoray, and S.~C. Glotzer.
\newblock {\em Nature}, 462:773--777, 2009.

\bibitem{Gardner1995}
M.~Gardner.
\newblock {\em New Mathematical Diversions (Revised Edition)}.
\newblock Math. Assoc. Amer., Washington, 1995.

\bibitem{Graaf2011}
J.~de~Graaf, R.~van Roij, and M.~Dijkstra.
\newblock {\em Phys. Rev. Lett.}, 105:155501, 2011.

\bibitem{Kallus2014}
Y.~Kallus.
\newblock {\em Adv. Math.}, 264:355--370, 2014.

\bibitem{Kallus2015}
Y.~Kallus.
\newblock {\em Geometry \& Topology}, 19(1):343--363, 2015.

\bibitem{Baule2013}
A.~Baule, R.~Mari, L.~Bo, L.~Portal, and H.~A. Makse.
\newblock {\em Nature Communications}, 4:2194, 2013.

\bibitem{Baule2014}
A.~Baule and H.~A. Makse.
\newblock {\em Soft Matter}, 10:4423--4429, 2014.

\bibitem{Philipse1996}
A.~P. Philipse.
\newblock {\em Langmuir}, 12(5):1127–1133, 1996.

\bibitem{Zong2014}
C.~Zong.
\newblock 2014.
\newblock arXiv:1410.1102.

\bibitem{Torquato2000}
S.~Torquato, T.~M. Truskett, and P.~G. Debenedetti.
\newblock {\em Phys. Rev. Lett.}, 84:2064, 2000.

\bibitem{Ohern2005}
N.~Xu, J.~Blawzdziewicz, and C.~S. O'Hern.
\newblock {\em Phys. Rev. E}, 71:061306, 2005.

\bibitem{convex-lipschitz}
F.~A. Toranzos.
\newblock {\em Amer. Math. Monthly}, 74(3):278–280, 1967.

\bibitem{Charbonneau2012}
P.~Charbonneau, E.~I. Corwin, G.~Parisi, and F.~Zamponi.
\newblock {\em Phys. Rev. Lett.}, 109:205501, 2012.

\bibitem{convolutions}
F.~E. Schuster.
\newblock {\em Transact. Amer. Math. Soc.}, 359:5567, 2007.

\bibitem{Jin2010}
Y.~Jin and H.~A. Makse.
\newblock {\em Physica A}, 389(23):5362--5379, 2010.

\bibitem{Note1}
The preprint source files include replication code for this calculation.

\bibitem{Jiao2010}
Y.~Jiao, F.~H. Stillinger, and S.~Torquato.
\newblock {\em Phys. Rev. E}, 81:041304, 2010.

\bibitem{Jaoshvili2010}
A.~Jaoshvili, A.~Esakia, M.~Porrati, and P.~M. Chaikin.
\newblock {\em Phys. Rev. Lett.}, 104(18):185501, 2010.

\bibitem{Kallus2011}
Y.~Kallus and V.~Elser.
\newblock {\em Phys. Rev. E}, 83:036703, 2011.

\end{thebibliography}

\appendix

\section{Appendix: definition of shape families}

In this appendix, we give definitions of the families
of shapes for which the slope of the random packing density curve was calculated
in \ref{tab:etas}.

An oblate ellipsoid of aspect ratio $\alpha$ is the region
$\{(x,y,z)\in\mathbb{R}^3:(x/\alpha)^2+(y/\alpha)^2+z^2\le 1\}$, where $\alpha>1$.
A prolate ellipsoid is the region
$\{(x,y,z)\in\mathbb{R}^3:x^2+y^2+(z/\alpha)^2\le 1\}$, where $\alpha>1$.
A lens is the intersection of two equal-sized spheres, and a spherocylinder is the convex hull of two equal-sized spheres.
A spherodisk is the convex hull of all equal-sized spheres with centers on a given circle,
and a spindle is the intersection of such a family of spheres. All shapes in these six 
families are axis symmetric.

We consider a few more bodies without axial symmetry. A
tetrahedral puff is the intersection of four equal-sized spheres with
centers at the corners of a regular tetrahedron \cite{Kallus2011}. 
A rounded tetrahedron is the convex hull of four such spheres.
Similarly, a triangular spindle and a rounded triangle, 
are, respectively, the intersection and convex hull of
three equal-sized spheres at the 
corners of an equilateral triangle.
A superball is the region of space determined by the inequality
$|x|^{2p}+|y|^{2p}+|z|^{2p}\le1$ \cite{Jiao2010}. When $p=1$, we recover the Euclidean ball.
When $1<p<\infty$, we call the superballs cubic, since they interpolate
between the ball and the cube. Similarly, superballs with $\tfrac12<p<1$ are called
octahedral, as they interpolate between the ball and the octahedron.

The infinitesimal deformations associated with these families are
\begin{gather}
    \begin{aligned}
	\rho_\text{obl-ell}(\mathbf{u}) &= -\rho_\text{prol-ell}(\mathbf{u}) = -u_z^2\\
	\rho_\text{lens}(\mathbf{u}) &= -\rho_\text{sph-cyl}(\mathbf{u}) = -|u_z|\\
	\rho_\text{sph-disk}(\mathbf{u}) &= -\rho_\text{spindle}(\mathbf{u}) = \sqrt{1-u_z^2}\\
	\rho_\text{rnd-tet}(\mathbf{u}) &= -\rho_\text{tet-puff}(\mathbf{u}) = \max_{i=1,2,3,4} \mathbf{u}\cdot\mathbf{v}^\text{tet}_i\\
	\rho_\text{rnd-tri}(\mathbf{u}) &= -\rho_\text{tri-spind}(\mathbf{u}) = \max_{i=1,2,3} \mathbf{u}\cdot\mathbf{v}^\text{tri}_i\\
	\rho_\text{cub-sup}(\mathbf{u}) &= -\rho_\text{oct-sup}(\mathbf{u})  = - \sum_{i=x,y,z} u_i^2\log(u_i)\text,
    \end{aligned}
\raisetag{5\baselineskip}\end{gather}
where $\mathbf{u}=(u_x,u_y,u_z)$ is a point on the unit sphere, $\mathbf{v}^\text{tet}_i$, $i=1,2,3,4$, are the four vertices
of a regular tetrahedron inscribed in the unit sphere, and $\mathbf{v}^\text{tri}_i$, $i=1,2,3$, are the three vertices
of an equilateral triangle inscribed in the equator.

\end{document}